\documentclass[sigconf]{acmart}

\def\BibTeX{{\rm B\kern-.05em{\sc i\kern-.025em b}\kern-.08emT\kern-.1667em\lower.7ex\hbox{E}\kern-.125emX}}

\usepackage{balance}

\usepackage{booktabs} 
\usepackage{multirow}
\usepackage{url}
\usepackage{paralist}
\usepackage{algorithm}
\usepackage{footnote}
\usepackage{threeparttable}
\usepackage{enumerate}
\usepackage{amssymb}
\usepackage{algpseudocode}

\newcommand\Algphase[1]{%
	\vspace*{-.4\baselineskip}\Statex\hspace*{\dimexpr-\algorithmicindent-2pt\relax}\rule{\columnwidth}{0.4pt}
	\Statex\hspace*{-\algorithmicindent}\textbf{#1}%
	\vspace*{-.6\baselineskip}\Statex\hspace*{\dimexpr-\algorithmicindent-2pt\relax}\rule{\columnwidth}{0.4pt}
}

\copyrightyear{2019}
\acmYear{2019}
\setcopyright{acmlicensed}
\acmConference[KDD '19] {The 25th ACM SIGKDD Conference on Knowledge Discovery and Data Mining}{August 4--8, 2019}{Anchorage, AK, USA}
\acmBooktitle{The 25th ACM SIGKDD Conference on Knowledge Discovery and Data Mining (KDD'19), June 22--24, 2019, Anchorage, AK, USA}
\acmPrice{15.00}
\acmDOI{10.1145/3292500.3330832}
\acmISBN{978-1-4503-6201-6/19/08}

\begin{document}

\fancyhead{}
\title{Exact-K Recommendation via Maximal Clique Optimization}

\author{
	Yu Gong$^{1,*}$,
	Yu Zhu$^{1,*}$,
	Lu Duan$^{2}$,
	Qingwen Liu$^{1}$,
	Ziyu Guan$^{3}$,
	Fei Sun$^{1}$,
	Wenwu Ou$^{1}$,
	Kenny Q. Zhu$^{4}$
}
\affiliation{
	\institution{$^{1}$ Alibaba Group, China\quad $^{2}$ Zhejiang Cainiao Supply Chain Management Co., Ltd, China}
	\institution{$^{3}$ Xidian University, China\quad $^{4}$ Shanghai Jiao Tong University, China}
}
\email{{gongyu.gy,zy143829,xiangsheng.lqw,ofey.sf}@alibaba-inc.com, duanlu.dl@cainiao.com}
\email{zyguan@xidian.edu.cn, santong.oww@taobao.com, kzhu@cs.sjtu.edu.cn}

\renewcommand{\authors}{Yu Gong, Yu Zhu, Lu Duan, Qingwen Liu, Ziyu Guan, Fei Sun, Wenwu Ou, Kenny Q. Zhu}
\renewcommand{\shortauthors}{Yu Gong et al.}

\begin{abstract}
This paper targets to a novel but practical recommendation problem named exact-K recommendation.
It is different from traditional top-K recommendation,
as it focuses more on (constrained) combinatorial optimization which will optimize to recommend a whole set of $K$ items called card,
rather than ranking optimization 
which assumes that ``better'' items should be put into top positions.
Thus we take the first step to give a formal problem definition,
and innovatively reduce it to Maximum Clique Optimization based on graph.
To tackle this specific combinatorial optimization problem which is NP-hard, we propose \emph{Graph Attention Networks} (GAttN) with a Multi-head Self-attention encoder and a decoder with attention mechanism.
It can end-to-end learn the joint distribution of the $K$ items and generate an optimal card rather than rank individual items by prediction scores.
Then we propose \emph{Reinforcement Learning from Demonstrations} (RLfD) which combines the advantages in behavior cloning and reinforcement learning, making it sufficient-and-efficient to train the model.
Extensive experiments on three datasets demonstrate the effectiveness of our proposed \emph{GAttN with RLfD} method, it outperforms several strong baselines with a relative improvement of 7.7\% and 4.7\% on average in Precision and Hit Ratio respectively, and achieves state-of-the-art (SOTA) performance for the exact-K recommendation problem.
\end{abstract}

%
%



\maketitle

\renewcommand{\thefootnote}{\fnsymbol{footnote}}
\footnotetext[1]{Equal contribution.}
\renewcommand{\thefootnote}{\arabic{footnote}}

\section{Introduction}
\label{sec:intro}

The explosive growth and variety of information (e.g. movies, commodities, news etc.) available on the web frequently overwhelms users,
while Recommender Systems (RS) are valuable means to cope with the information overload problem. 
RS usually provide the target user with a list of items,
which are selected from the overwhelmed candidates
in order to best satisfy his/her current demand.
In the most traditional scenarios of RS especially on mobiles,
recommended items are shown in a waterfall flow form,
i.e. users should scroll the screen and items will be presented one-by-one.
Due to the pressure of QPS (Query-Per-Second) for users interacting with RS servers, it is common to return a large amount of ranked items
(e.g. 50 items in Taobao RS)
based on CTR (Click-Through-Rate) estimation for example\footnote{Here we take CTR as an example, other preference score can also be used, e.g. Movie Rating, CVR (Conversion-Rate) or GMV (Gross-Merchandise-Volume) etc.} and present them from top to bottom.
That is to say we believe the top ranked items take the most chance to be clicked or preferred so that when users scroll the screen and see items top-down,
the overall clicking efficiency can be optimized.
It can be seen as \emph{top-K} recommendation \cite{cremonesi2010performance}, because the ranking of item list is important.
\begin{figure*}[th]
	\centering
	\includegraphics[angle=0, width=1.9\columnwidth]{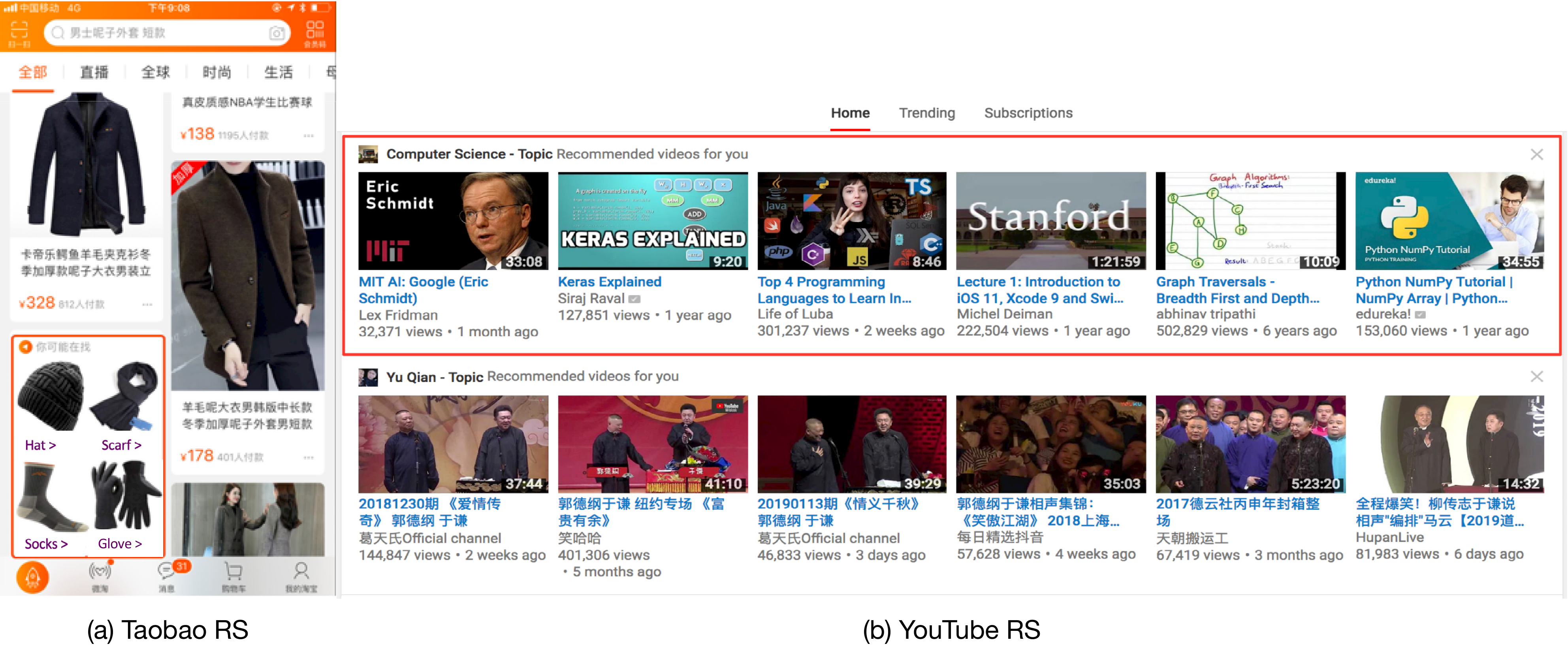}
	\caption{Show cases for exact-K recommendation in Taobao and YouTube.}
	\label{fig:exactk_demo}
\end{figure*}

However in many real-world recommendation applications,
exact $K$ items are shown once all to the users. 
In other words, users should not scroll the screen and the combination of $K$ items is shown as a whole \textbf{card}. 
Taking two popular RS in the homepages of Taobao and YouTube for example
(illustrated in Fig. \ref{fig:exactk_demo}), they recommend cards with exact 4 commodities and 6 videos respectively.
Note that items in the same card may interact with each other, e.g. in Taobao, co-occurrence of ``hat'' and ``scarf'' performs better than ``shoe'' and ``scarf'', but ``shoe'' and ``scarf'' can be optimal individually.
We call it \emph{exact-K} recommendation, 
whose key challenge is to maximize the chance of the whole card being clicked or satisfied by the target user. 
Meanwhile, items in a card usually maintain some \textbf{constraints} between each other to guarantee the user experience in RS,
e.g. the recommended commodities in E-commerce should have some diversity rather than being all similar for complement consideration.
In a word, top-K recommendation can be seen as a ranking optimization problem which assumes that ``better'' items should be put into top positions,
while exact-K recommendation is a (constrained) combinatorial optimization problem
which tries to maximize the joint probability of the set of items in a card.

Top-K recommendation has been well studied for decades in information retrieval (IR) research community. Among them, listwise models are the most related to our problem as they also perform optimization considering the whole item list. However, they either target on ranking refinement or do not consider constraints in the ranking list, which will fall into sub-optimal towards exact-K recommendation (refer to Sec. \ref{sec:ltr} for more discussions). Our work mainly focuses on solving exact-K recommendation problem end-to-end, and its main contributions can be summarized as follows.
\begin{enumerate}[(1)]
\item
We take the first step to formally define the exact-K recommendation problem and innovatively reduce it to a Maximal Clique Optimization problem based on graph.
\item
To solve it, we propose \emph{Graph Attention Networks} (GAttN) with an Encoder-Decoder framework which can end-to-end learn the joint distribution of $K$ items and generate an optimal card containing $K$ items.
Encoder utilizes Multi-head Self-attention to encode the constructed undirected graph into node embeddings considering nodes correlations. Based on the node embeddings, decoder generates a clique consisting of $K$ items with RNN and attention mechanism which can well capture the combinational characteristic of the $K$ items.
Beam search with masking is applied to meet the constraints.
Then we adopt well-designed \emph{Reinforcement Learning from Demonstrations} (RLfD) which combines the advantages in behavior cloning and reinforcement learning, making it sufficient-and-efficient to train GAttN.
\item We conduct extensive experiments on three datasets (two constructed from public MovieLens datasets and one collected from Taobao).
Both quantitative and qualitative analysis justify the effectiveness and rationality of our proposed \emph{GAttN with RLfD} for exact-K recommendation.
Specifically, our method outperforms several strong baselines with significant improvements of 7.7\% and 4.7\% on average in Precision and Hit Ratio respectively.
\end{enumerate}

\section{Related Works}
\subsection{Top-K Recommendation}
\label{sec:top_k_related}
Top-K recommendation refers to recommending a list of $K$ ranked items to a user, which is related to the descriptions of recommendation problem and learning to rank methods.
\subsubsection{The Recommendation Problem}
The key problem of recommendation system lies in how to generate users' most preferred item list.
Some previous works \cite{koren2009matrix} model the recommendation problem as a regression task (i.e. predict users' ratings on items) or classification task (i.e. predict whether the user will click/purchase/\ldots the item).
Items are then ranked based on the regression scores or classification probabilities to form the recommendation list.
Other works \cite{rendle2009bpr,wang2017irgan,he2017neural} directly model the recommendation problem as a ranking task, where many pairwise/listwise ranking methods are exploited to generate users' top-k preferred items.
Learning to rank is surveyed in detail in the next subsection.

\subsubsection{Learning to Rank}
\label{sec:ltr}
Learning to Rank (LTR) refers to a group of techniques that attempts to solve ranking problems by using machine learning algorithms.
It can be broadly classified into three categories: pointwise, pairwise, and listwise models.
Pointwise Models \cite{koren2009matrix,he2017neural} treat the ranking task as a classification or regression task.
However, pointwise models do not consider the inter-dependency among instances in the final ranked list.
Pairwise Models \cite{rendle2009bpr} assume that the relative order between two instances is known and transform it to a pairwise classification task.
Note that their loss functions only consider the relative order between two instances,
while the position of instances in the final ranked list can hardly be derived.
Listwise Models provide the opportunity to directly optimize ranking criteria and achieve whole-page ranking optimization.
Recently \cite{ai2018learning} proposed Deep Listwise Context Model (DLCM) to fine-tune the initial ranked list generated by a base model,
which achieves SOTA performance.
Other whole-page ranking optimization methods can be found in \cite{jiang2018beyond},
which mainly focus on ranking refinement.
Listwise models are the most related to our problem.
However, they either target on ranking refinement or don't consider the constraints in ranking list,
which are not well-designed for exact-K recommendation.

\subsection{Neural Combinatorial Optimization}
\label{sec:nco_related}
Even though machine learning (ML) and combinatorial optimization have been studied for decades respectively,
there are few investigations on the application of ML methods in solving the combinational optimization problem.
Current related works mainly focus on two types of ML methods: supervised learning and reinforcement learning.
Supervised learning \cite{vinyals2015pointer} is the first successful attempt to solve the combinatorial optimization problem.
It proposes a special attention mechanism named Pointer-Net to tackle a classical combinational optimization problem: Traveling Salesman Problem (TSP).
Reinforcement learning (RL) aims to transform the combinatorial optimization problem into a sequential decision problem
and becomes increasingly popular recently.
Based on Pointer Network, \cite{bello2016neural} develops a neural combinatorial optimization framework with RL, which performs excellently in some classical problems, such as TSP and Knapsack Problem.
RL is also applied to
RS \cite{zheng2018drn},
but it is still designed for traditional top-K recommendation.
In this work, we focus on exact-K recommendation,
which is transferred into the maximal clique optimization problem.
Some researches \cite{Ion2011Image} also try to solve it,
but they often focus on estimation of node-weight.
The main difference between them and ours is that we target to directly select an optimal clique rather than search for the clique comprised of maximum weighted nodes, which brings a grave challenge.
\section{Problem Definition}
In this section, we first give a formal definition of exact-K recommendation, and then discuss how to transfer it to the Maximal Clique Optimization problem. Finally we provide a baseline approach to tackle the above problem.
\subsection{Exact-K Recommendation}
\label{sec:problem_definition}
Given a set of candidate $N$ items $S=\{s_i\}_{1\le i\le N}$,
our goal is to recommend exact $K$ items $A=\{a_i\}_{1\le i\le K}\subseteq S$ which is shown as a whole card\footnote{Here we suppose that permutation of the $K$ items in a card is not considered in exact-K recommendation.},
so that the combination of items $A$ takes the most chance to be clicked or satisfied by a user $u$.
We denote the probability of $A$ being clicked/satisfied as $P(A,r=1|S,u)$.
Somehow items in $A$ should obey some $M$ constraints between each other as $C=\{c_k(a_i,a_j)=1|a_i\in A,a_j\in A,i\neq j\}_{1\le k\le M}$ or not as $C=\emptyset$, here $c_k$ is a boolean indicator
which will be $1$ if the two items satisfy the constraint.
Overall the problem of exact-K recommendation can be regarded as a (constrained) combinatorial optimization problem,
and is defined formally as follows:
\begin{small}
\begin{eqnarray}
	\label{eq:objective}
	& \max\limits_{A} P(A,r=1|S,u;\theta), \\
	\label{eq:constraints}
	& s.t\ \forall a_i\in A,a_j\in A,i\neq j, \forall c_k\in C, c_{k}(a_i,a_j)=1,
\end{eqnarray}
\end{small}
where $\theta$ is the parameters for function of generating $A$ from $S$ given user $u$, and $r=1$ donates relevance/preference indicator.

\begin{figure}[th]
	\centering
	\includegraphics[angle=0, width=0.81\columnwidth]{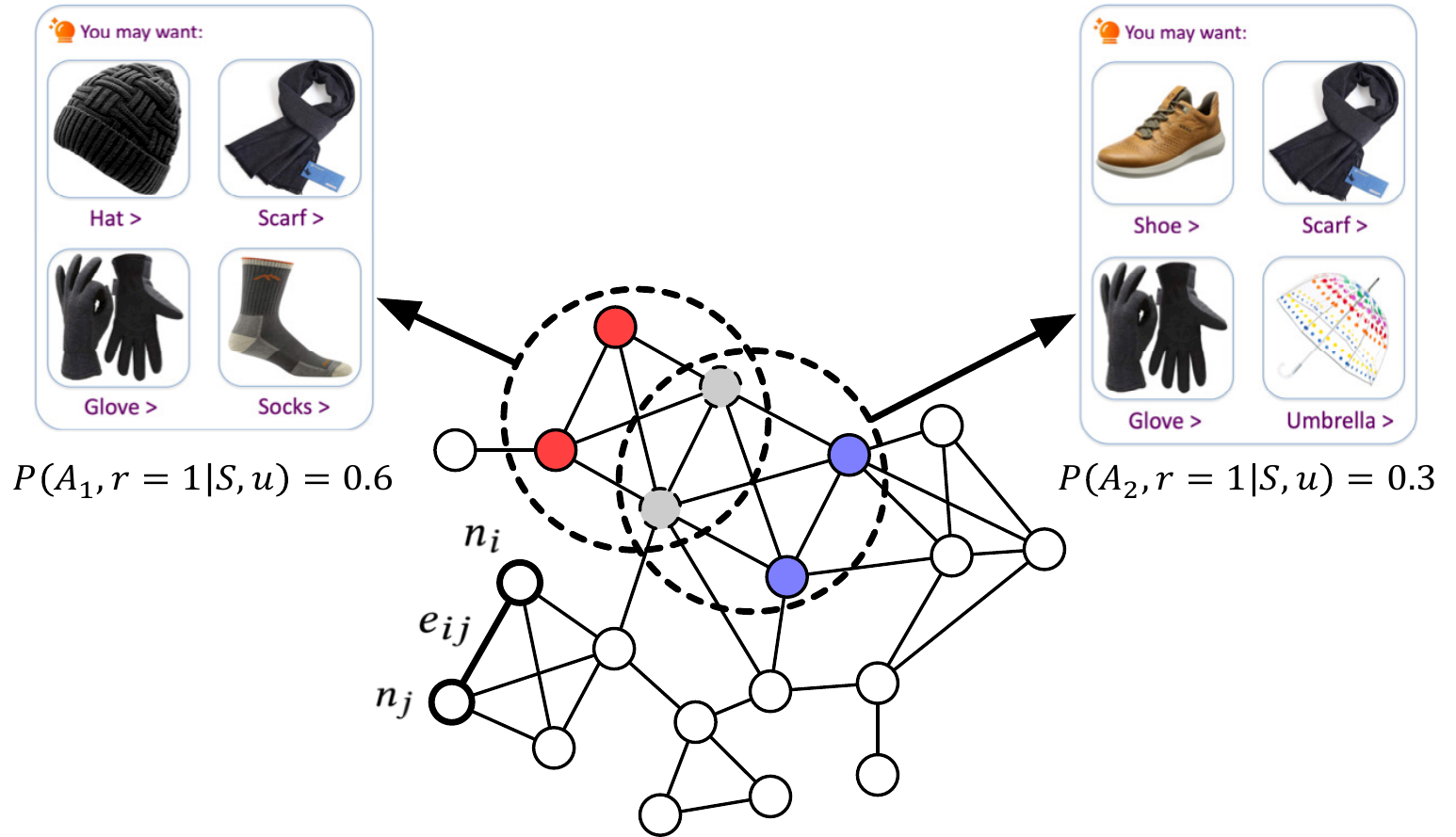}
	\caption{Illustration for a specific graph $\mathbb{G}$ with $N=20$ and $K=4$.
	We show two different cliques (red and blue) in graph and the corresponding cards ($A_1$ and $A_2$) each with 4 items.
	We suppose that $P(A_1,r=1|S,u)>P(A_2,r=1|S,u)$ means that card $A_1$ takes more chance to be satisfied than card $A_2$ given user $u$ and candidate item set $S$.}
	\label{fig:clique}
\end{figure}

In another perspective, we construct a graph $\mathbb{G}(\mathcal{N},\mathcal{E})$ containing $N$ nodes, in which each node $n_i$ in $\mathcal{N}$ represents an item $s_i$ in candidate item set $S$, each edge $e_{ij}$ in $\mathcal{E}$ connecting nodes $(n_i,n_j)$ represents that items $s_i$ and $s_j$ should satisfy the constraints or there is no constraint (a.k.a $\mathbb{G}$ is now a complete graph),
i.e $\forall c_k\in C, c_{k}(s_i,s_j)=1$ or $C=\emptyset$, so it is an undirected graph here.
Intuitively, we can transfer exact-K recommendation into the maximal clique optimization problem\footnote{It can be generalized according to the optimization objective.} \cite{garey1974some,gong2016representing}. 
That is to say we aim to select a clique\footnote{A clique is a subset of nodes of an undirected graph such that every two distinct nodes in the clique are adjacent; that is, its induced subgraph is complete.}
(i.e characteristics of clique can ensure the constraints defined in Eq. \ref{eq:constraints})
with $K$ nodes from $\mathbb{G}$ where combination of the selected $K$ corresponding items $A$ achieves the maximal objective defined in Eq. \ref{eq:objective}.
You can take Fig. \ref{fig:clique} as an example.
Furthermore, maximal clique problem is proved to be NP-hard thus it can not be solved in polynomial time \cite{garey1974some}.

\subsection{Naive Node-Weight Estimation Method}
\label{sec:nnwem}
A baseline method is that we can estimate a weight as $w_i$ of each node $n_i\in \mathcal{N}$ related to the optimization objective in graph $\mathbb{G}$.
In exact-K recommendation, our goal is to maximize the clicked or satisfied probability of the recommended $K$ items set as in Eq. \ref{eq:objective},
so we regard the weight of each node as CTR of corresponding item.
After getting the weight of each node in graph supported as $\mathbb{G}(\mathcal{N},\mathcal{E},\mathcal{W})$,
we can reduce the Maximal Clique Optimization problem as  
finding a clique in graph $\mathbb{G}$ with maximal node weights summation.
We can then apply some heuristic methods like Greedy search to solve it.
Specifically, we modify Eq. \ref{eq:objective} as follows:
\begin{small}
\begin{equation}
	\max\limits_{A} \sum_{a_i\in A}{P(r=1|a_i,S,u;\theta)},
\end{equation}
\end{small}
where $P(r=1|a_i,S,u;\theta)$ can be regarded as node weight $w_i$ in graph.
Here we focus on how to estimate the node weights $\mathcal{W}$,
it can be formulated as a normal item CTR estimation task in IR.
A large amount of LTR based methods for CTR estimation can be adopted as our strong baselines. Refer to Sec. \ref{sec:ltr} for more details. 

We call the adapted baseline as Naive Node-Weight Estimation Method, with its detailed implementation shown in Algorithm \ref{alg:naive}~\footnote{In our problem, we ignore the circumstance of getting infeasible solution, and we argue that in real-world application with small $K$ and large $N$ we can always find a clique with $K$ nodes in graph with $N$ nodes greedily.}.
However weaknesses of such method are obvious for the following three points:
\begin{inparaenum}
\item CTR estimation for each item is independent, 
\item combinational characteristic of the $K$ items in a card is not considered,
\item problem objective is not optimized directly but substituted with a reduced heuristic objective which will unfortunately fall into sub-optimal.
\end{inparaenum}
On the contrary, we will utilize a framework of Neural Combinatorial Optimization (some related works in Sec. \ref{sec:nco_related}) to directly optimize the problem objective in Sec. \ref{sec:approach}.

\section{Approach}
\label{sec:approach}
\begin{figure*}[th]
	\centering
	\includegraphics[angle=0, width=1.75\columnwidth]{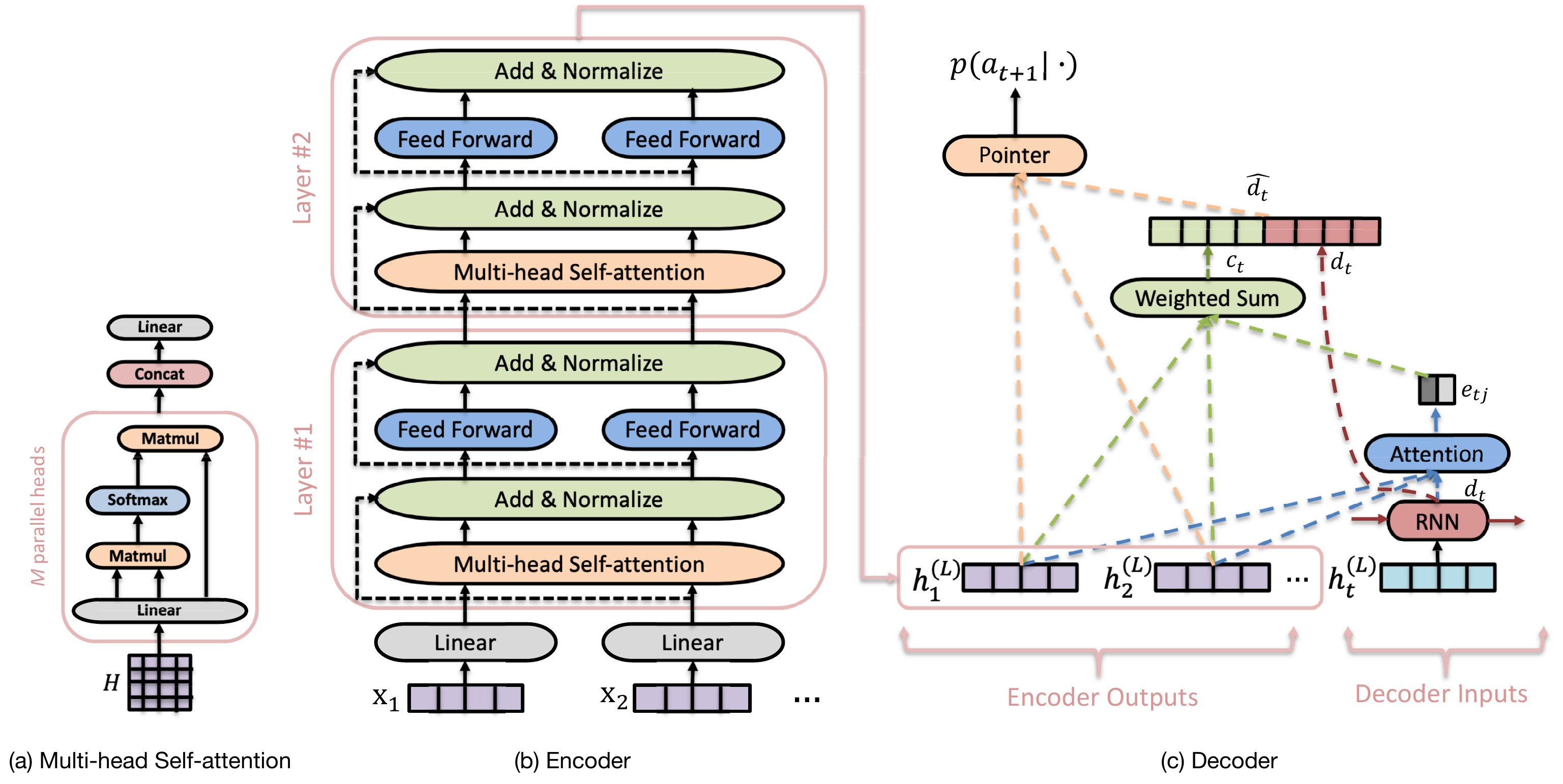}
	\caption{The key modules of Graph Attention Networks (GAttN).}
	\label{fig:transformer}
\end{figure*}
In Sec. \ref{sec:problem_definition}, we formally define the exact-K recommendation problem based on searching for maximal scoring clique with $K$ nodes in a
specially constructed graph $\mathbb{G}(\mathcal{N},\mathcal{E})$ with $N$ nodes. 
The score of a clique is the overall probability of a user clicking or stratifying the corresponding card of $K$ items as in Eq. \ref{eq:objective}.
To tackle this specific problem, we first propose \emph{Graph Attention Networks} (GAttN) 
which follows the Encoder-Decoder Pointer framework \cite{vinyals2015pointer} with Attention mechanism \cite{vaswani2017attention,bahdanau2014neural}.
Then we adopt \emph{Reinforcement Learning from Demonstrations} (RLfD) which combines the advantages in behavior cloning \cite{torabi2018behavioral} and reinforcement learning, making it sufficient-and-efficient to train the networks.
\subsection{Graph Attention Networks}
\label{sec:gattn}
The traditional encoder-decoder framework \cite{sutskever2014sequence} usually encodes an input sequence into a fixed vector by a recurrent neural networks (RNN)
and decodes a new output sequence from that vector using another RNN. 
For our exact-K recommendation problem, the encoder produces representations of all input nodes in graph $\mathbb{G}$, 
and the decoder selects a clique $A$ among input nodes by pointer, in which the constraint of clique is considered by masking.

\subsubsection{Input}
\label{sec:input}
We first define the input representation of each node in graph $\mathbb{G}(\mathcal{N},\mathcal{E})$.
Specifically in our problem, given candidate items set $S$ and user $u$,
we can represent the input $x_i$ of a node $n_i\in \mathcal{N}$ by combination of the features of corresponding item $s_i\in S$ and user $u$.
Here we use a simple fully connected neural network with nonlinear activation ReLU as:
\begin{small}
\begin{equation}
	x_i=ReLU(W_I[x_{s_i};x_u]+b_I),
\end{equation}
\end{small}
where $x_{s_i}$ and $x_u$ are feature vectors for item $s_i$ and user $u$ (e.g trainable embeddings of corresponding item and user IDs), $[a;b]$ represents the concatenation of vector $a$ and $b$, $W_I$ and $b_I$ are training parameters for input representation.

\subsubsection{Encoder}
\label{sec:encoder}
First of all, since the order of nodes in a undirected graph is meaningless,
the encoder network architecture should be permutation invariant,
i.e. any permutation of the inputs results in the same output representations. 
While the traditional encoder usually uses a RNN to convey sequential information, e.g., in text translation the relative position of words must be captured,
but it is not appropriate to our case.
Secondly, the representation for a node should consider the other nodes in graph, as there can exist some underlying structures in graph that nodes may influence between each other.
So it's helpful to represent a node with some attentions to other nodes.
As a result, we use a model like Self-attention,
it is a special case of attention mechanism that only requires a single sequence to compute its representation.
Self-attention has been successfully applied to many NLP tasks up to now \cite{yang2018query}, here we utilize it to encode the graph and produce nodes representations. 

Actually in this paper, the encoder that we use is similar
to the encoder used in Transformer architecture 
by \cite{vaswani2017attention} with multi-head self-attention.
Fig. \ref{fig:transformer}(b) depicts the computation graph of encoder.
From the $d_x$-dimensional input feature vector $x_i$ for node $n_i$, the encoder firstly computes initial $d_h$-dimensional graph node embedding (a.k.a representation) $h_i^{(0)}$ through a learned linear projection with parameters $W_E$ and $b_E$ as:
\begin{small}
\begin{equation}
	h_i^{(0)} = W_Ex_i + b_E.
\end{equation}
\end{small}
The node embeddings are then updated through $L$ self-attention layers, each consisting of two sub-layers:
a multi-head self-attention (MHSA) layer followed by a feed-forward (FF) layer.
We denote with $H^{(l)}=\{h^{(l)}_i\}_{1\leq i\leq N}$ the graph node embeddings produced by layer $l$,
and the final output graph node embeddings as $H^{(L)}$.

The basic component of MHSA is the scaled dot-product attention, which is a variant of dot-product (multiplicative) attention \cite{luong2015effective}.
Given a matrix of $N$ input $d$-dimensional embedding vectors $E\in \mathbb{R}^{N\times d}$,
the scaled dot-product attention computes the self-attention scores based on the following equation:
\begin{small}
\begin{equation}
\text{SelfAttention}(E)=softmax(\frac{EE^T}{\sqrt{d}})E,
\end{equation}
\end{small}
where $softmax$ is row-wise. 
More specifically,
MHSA sub-layers will employ $M$ attention heads to capture multiple attention information
and the results from each head are concatenated followed by a parameterized linear transformation to produce the sub-layer outputs.
Fig. \ref{fig:transformer}(a) shows the computation graph of MHSA.
Specifically in layer $l$, it will operate on the output embedding matrices
$H^{(l-1)}\in \mathbb{R}^{N\times d_h}$ from previous layer $l-1$
and produce the MHSA sub-layer outputs $\widehat{H^{(l)}}\in \mathbb{R}^{N\times d_h}$ as:
\begin{small}
\begin{equation}
\begin{aligned}
&\widehat{H^{(l)}}=[head_1;\dots;head_M]W_O, \\
\text{where}\ &head_i=\text{SelfAttention}(H^{(l-1)}W_{hi}),
\end{aligned}
\end{equation}
\end{small}
where $M$ is the number of heads,
$W_{hi}\in \mathbb{R}^{d_h\times d_k}$ is parameter for each head,
$W_O\in \mathbb{R}^{(Md_k)\times d_h}$ is parameter for linear transformation output,
and $d_k$ is the output dimension in each head.
In addition to MHSA sub-layers, FF sub-layers consist of two linear transformations with a ReLU activation in between.
\begin{small}
\begin{equation}
h_i^{(l)} = W_{F2}ReLU(W_{F1}\widehat{h_i^{(l)}}+b_{F1})+b_{F2},
\end{equation}
\end{small}
where $W_{F1}, W_{F2}$ are parameter matrices, $\widehat{h_i^{(l)}}$ and $h_i^{(l)}$ represent embedding outputs of node $n_i$ in MHSA and FF sub-layers correspondingly.
We emphasize that those trainable parameters mentioned above are unique per layer.
Furthermore, each sub-layer adds a skip-connection \cite{he2016deep} and layer normalization \cite{ba2016layer}.

As a result, \emph{encoder} transforms $x_1,x_2,\cdots,x_N$ (original representations of nodes in graph) in Fig. \ref{fig:transformer}(b)
to $h_1^{(L)},h_2^{(L)},\cdots,h_N^{(L)}$ (embedding representations of these nodes, considering graph constructure information), which will be used in decoder in Fig. \ref{fig:transformer}(c).

\subsubsection{Decoder}
\label{sec:decoder}
For exact-K recommendation problem, the output $A$ represents a clique (card) with $K$ nodes (items) in graph $\mathbb{G}$ that can be interrelated with each other.
Recently, RNN \cite{vinyals2015pointer,gong2018deep,gong2018automatic} has been widely used to map the encoder embeddings to a correlated output sequence,
so does in our proposed framework. 
We call this RNN architecture decoder to decode the output nodes $A=\{a_1,\dots,a_K\}$.
Remember our goal is to optimize $P(A|S,u;\theta)$ defined in Sec. \ref{sec:problem_definition} (here we omit relevance score of $r=1$),
it is a joint probability and can be decomposed by the chain rule as follows:
\begin{small}
\begin{equation}
\label{eq:chain_rule}
\begin{aligned}
P(A|f(S,u;\theta)&=\prod_{i=1}^{K}p(a_i|a_1,\dots,a_{i-1},S,u;\theta)\\
&=\prod_{i=1}^{K}p(a_i|a_1,\dots,a_{i-1},f(S,u;\theta_e);\theta_d),
\end{aligned}
\end{equation}
\end{small}
where we represent encoder as $f(S,u;\theta_e)$ with trainable parameters $\theta_e$, and decoder trainable parameters as $\theta_d$.
The last term in above
Eq. \ref{eq:chain_rule} is estimated with RNN by introducing a state vector,
$d_i$, which is a function of the previous state $d_{i-1}$, and the previous output node $a_{i-1}$, i.e.
\begin{small}
\begin{equation}
p(a_i|a_1,\cdots,a_{i-1},f(S,u;\theta_e);\theta_d)=p(a_i|d_{i},f(S,u;\theta_e);\theta_d),
\end{equation}
\end{small}
where $d_i$ is computed as follows:
\begin{small}
\begin{eqnarray}
\label{eq:rnn_hidden}
d_i = 
\begin{cases}
g(0,0) & \text{if}\ i=1, \\
g(d_{i-1},a_{i-1}) & \text{otherwise},
\end{cases}
\end{eqnarray}
\end{small}
where $g(d,a)$ is usually a non-linear function (e.g. cell in LSTM \cite{zhu2017next} or GRU \cite{zhu2018brand}) that combines the previous state and previous output
(embedding of the corresponding node $a$ from encoder) in order to produce the current state.

Decoding happens sequentially, 
and at time-step $t \in \{1,\dots,K\}$, 
the decoder outputs the node $a_t$ based on the output embeddings from encoder
and already generated outputs $\{a_{t'}\}_{1\leq t'\leq t}$ which are embedded by RNN hidden state $d_t$. 
See Fig. \ref{fig:transformer}(c) for an illustration of the decoding process.
During decoding, 
$p(a_t|d_t,f(S,u;\theta_e);\theta_d)$ is implemented by an specific attention mechanism named \textbf{pointer} \cite{vinyals2015pointer},
in which it will attend to each node in the encoded graph 
and calculate the attention scores before applying \emph{softmax} function to get the probability distribution.
It allows the decoder to look at the whole input graph $\mathbb{G}(\mathcal{N},\mathcal{E})$ at any time
and select a member of the input nodes $\mathcal{N}$ as the final outputs $A$. 

For notation purposes, let us define decoder output hidden states as $(d_1,\dots,d_K)$,
the encoder output graph node embeddings as $(h_1^{(L)},\dots,h_N^{(L)})$.
At time-step $t$, decoder first glimpses \cite{vinyals2015order} the whole encoder outputs,
and then computes the representation of decoding up to now together with attention to the encoded graph, denoted as $\widehat{d_t}$ and the equation is as follows:
\begin{small}
\begin{eqnarray}
\label{eq:glimpse}
&e_{tj}=softmax(v_{D1}^T\tanh(W_{D1}d_t+W_{D2}h_j^{(L)})),j\in\{1,\dots,N\}, \nonumber\\
&c_t=\sum_{j=1}^{N}e_{tj}h_j^{(L)},\ \widehat{d_t}=[d_t;c_t],
\end{eqnarray}
\end{small}
where $W_{D1}$, $W_{D2}$ and $v_{D1}$ are trainable parameters.
After getting the representation of decoder at time-step $t$,
we apply a specific attentive pointer with masking scheme to generate feasible clique from graph.
In our case, we use the following masking procedures: 
\begin{inparaenum}
\item nodes already decoded are not allowed to be pointed,
\item nodes will be masked if disobey the clique constraint rule among the decoded subgraph.
\end{inparaenum}
And we compute the attention values as follows:
\begin{small}
\begin{equation}
\label{eq:pointer}
u_{tj}=
\begin{cases}
v_{D2}^T\tanh{(W_{D3}\widehat{d_t}+W_{D4}h_j^{(L)})},\ \ \text{otherwise},  \\
-\infty,\ \  \text{if node $n_j$ should be masked,}
\end{cases} \\
\end{equation}
\end{small}
where $v_{D2}$, $W_{D3}$, and $W_{D4}$ are trainable parameters.
Then \emph{softmax} function is applied to get the pointing distribution towards input nodes, as follows:
\begin{small}
\begin{equation}
p(a_t|d_t,f(S,u;\theta_e);\theta_d)=softmax(u_{tj}),\ j\in\{1,\dots,N\}.
\end{equation}
\end{small}
We mention that the attention mechanism adopted in Eq. \ref{eq:glimpse} and \ref{eq:pointer} is following Bahdanau et al \cite{bahdanau2014neural}.
At the period of decoder inference, we apply technique of beam search \cite{vinyals2015show}.
It is proposed to expand the search space and try more combinations of nodes in a clique (a.k.a items for a card) to get a most optimal solution.

To summarize, \emph{decoder} receives embedding representations of nodes in graph $\mathbb{G}$ from encoder, and selects clique $A$ of $K$ nodes with attention mechanism.
With the help of RNN and beam search, decoder in our proposed GAttN framework is able to capture the combinational characteristics of the $K$ items in a card. 

\subsection{Reinforcement Learning from Demonstrations}
\label{sec:rlfd}
\subsubsection{Overall}
In our proposed GAttN framework,
we represent encoder as $f(S,u;\theta_e)$ which can be seen as \textbf{state} $\mathcal{S}$ in RL,
and we represent decoder as $P(A|f(S,u;\theta_e);\theta_d)=P(A|\mathcal{S};\theta_d)$ which can be seen as \textbf{policy} $\mathcal{P}$ in RL. 
A \emph{Reinforcement Learning from Demonstration} (RLfd) agent, possessing
both an objective reward signal and demonstrations, is able to
combine the best from both fields.
This framework is first proposed in domain of Robotics \cite{nair2018overcoming}.
Learning from demonstrations is much sample efficient and can speed up learning process,
leveraging demonstrations to
direct what would otherwise be uniform random exploration and
thus speed up learning. 
While the demonstration trajectories may be noisy or sub-optimal,
so policy supervised from such demonstrations will be worse too.
And learning from demonstrations is not directly targeting the objective which makes the policy fall into local-minimal.
On the other hand,
training policy by reinforcement learning can directly optimize the objective reward signal,
witch allows such an agent to eventually outperform a
potentially sub-optimal demonstrator.

\subsubsection{Learning from Demonstrations}
\label{sec:supervise}
Learning from demonstrations can be seen as behavior cloning imitation learning \cite{torabi2018behavioral},
it applies supervised learning for policy (mapping states to actions)
using the demonstration trajectories as ground-truth.
We collect the ground truth clicked/satisfied cards $A^*=\{a_i^*\}_{1\leq i\leq K}$ given user $u$ and candidate items set $S$ as
demonstration trajectories and formulated as $P_{data}^{S}(A^*|S,u)$,
we can define the following loss function based on cross entropy $CrossEntropy$ of the generated cards $P(A|S,u;\theta)$ and demonstrated cards $P^S_{data}(A^*|S,u)$.
\begin{small}
\begin{equation}
\begin{aligned}
\mathcal{L}_{S}(\theta)&=\sum_{S,u}{CrossEntropy(P(A|S,u;\theta),P_{data}^{S}(A^*|S,u))} \\
&= -\sum_{S,u,A^*\in P_{data}^{S}}\log{P(A^*|S,u;\theta)} \\
&= -\sum_{S,u,A^*\in P_{data}^{S}}\sum_{i=1}^{K}\log{p(a_i^*|a_1^*,\cdots,a_{i-1}^*,\mathcal{S};\theta_d)} \\
&= -\sum_{S,u,A^*\in P_{data}^{S}}\sum_{i=1}^{K}\log{p(a_i^*|d_i^*,\mathcal{S};\theta_d)},
\end{aligned}
\label{eq:cross_entropy}
\end{equation}
\end{small}
where $d_i^*$ in the last term is state vector estimated by a RNN defined in Eq. \ref{eq:rnn_hidden} with inputs of $d_{i-1}^*$ and $a_{i-1}^*$.
This means that the decoder model focuses on learning to output the next item of the card given the current state of the model
AND previous ground-truth items.

\paragraph{SUPERVISE with Policy-sampling.}
During inference the model can generate a full card $A$ given state $\mathcal{S}$ by generating one item at a time until we get $K$ items.
For this process, at time-step $t$, the model needs the output item $a_{t-1}$ from the last time-step as input in order to produce $a_t$.
Since we don't have access to the true previous item, we can instead either select the most likely one given our model or sample according to it.
In all these cases, if a wrong decision is taken at time-step $t-1$, the model can be in a part of the state space that is very different from those visited from the training distribution and for which it doesn't know what to do.
Worse, it can easily lead to cumulative bad decisions.
We call this problem as discrepancy between training and inference \cite{bengio2015scheduled}.

In our work, we propose \emph{SUPERVISE with Policy-sampling} to bridge the gap between training and inference of policy.
We change the training process from fully guided using the true previous item,
towards using the generated item from trained policy instead.
The loss function for \emph{Learning from Demonstrations} is now as follows:
\begin{small}
\begin{equation}
\label{eq:L_S}
\begin{aligned}
\mathcal{L}_{S}(\theta)&=-\sum_{S,u,A^*\in P_{data}^{S}}\sum_{i=1}^{K}\log{p(a_i^*|a_1,\cdots,a_{i-1},\mathcal{S};\theta_d)} \\
&=-\sum_{S,u,A^*\in P_{data}^{S}}\sum_{i=1}^{K}\log{p(a_i^*|d_{i},\mathcal{S};\theta_d)},
\end{aligned}
\end{equation}
\end{small}
where $d_i$ is computed by Eq. \ref{eq:rnn_hidden} with inputs of $d_{i-1}$ and $a_{i-1}$ now, here $a_{i-1}$ is sampled from the trained policy $p(a_{i-1}|d_{i-1},\mathcal{S};\theta_d)$.

\subsubsection{Learning from Rewards}
\label{sec:reinforce}
\paragraph{Reward Estimator.}
The objective of exact-K recommendation is to maximize the chance of being clicked or satisfied for the selected card $A$ given candidate items set $S$ and user $u$,
as we defined in Sec. \ref{sec:problem_definition} and Eq. \ref{eq:objective}.
Leveraging the advantage of reinforcement learning, we can directly optimize the objective by regarding it as \textbf{reward} function in RL.
While there is no explicit reward in our problem, 
we can estimate the reward function based on teacher's demonstration by the idea from Inverse Reinforcement Learning \cite{abbeel2004apprenticeship}.
The reward function can then be more generalized against supervised by demonstration only.
In our problem, there are large amount of explicit feedback data in which users click cards (labeled as $r^*=1$) or not (labeled as $r^*=0$),
we represent it as $P_{data}^{D}(r^*|A,u)$.
Then we transfer estimation of reward function to the problem of CTR estimation for a card $A$ given user $u$ as $P(r=1|A,u;\phi)$,
and the loss function for training it is as follows:
\begin{small}
\begin{eqnarray}
	& \mathcal{L}_{D}(\phi)=-\sum_{A,u,r^*\in P_{data}^{D}}&\Big(r^*\log\big(P(r=1|A,u;\phi)\big)\ + \\
	& & (1-r^*)\log\big(1-P(r=1|A,u;\phi)\big)\Big). \nonumber
\end{eqnarray}
\end{small}
To model the reward function, we follow the work of PNN \cite{qu2016product}
, in which we consider the feature crosses for card of items and user.
And we define $P(r=1|A,u;\phi)$ as following equation:
\begin{small}
\begin{eqnarray}
	& P(r=1|A,u;\phi) =  \\
	& \sigma\Big(W_{R2}ReLU\big(W_{R1}\big[[x_{a_i}\odot x_{u}]_{i=1}^{K};[x_{a_i}]_{i=1}^{K};x_{u}\big]+b_{R1}\big)+b_{R2}\Big), \nonumber
\end{eqnarray}
\end{small}
where $[\cdot]_{i=1}^{K}$ represents the concatenation of $K$ vectors, $\odot$ is inner-product and $\sigma$ means sigmoid function,
$x_{a_i}$ and $x_{u}$ are feature vector for item $a_i$ and user $u$ defined in Sec. \ref{sec:input},
$W_{R1},W_{R2},b_{R1},b_{R2}$ are trainable parameters for reward function totally donated by $\phi$.

\paragraph{REINFORCE with Hill-climbing.}
After we get the optimized reward function represented as $P(r=1|A,u;\phi^*)$,
we use policy gradient based reinforcement learning (REINFORCE) \cite{wang2017irgan} to train the policy.
And its loss function given previously defined dataset $P_{data}^{S}(\cdot|S,u)$ is derived as follows:
\begin{small}
\begin{eqnarray}
\label{eq:L_R}
&\mathcal{L}_{R}(\theta) & = -\sum_{S,u\in P_{data}^{S}}\mathbb{E}_{A\sim P(A|S,u;\theta)}[R(A,u)] \\
& & = -\sum_{S,u\in P_{data}^{S}}R(A,u)\sum_{i=1}^{K}\log{p(a_i|a_1,\cdots,a_{i-1},\mathcal{S};\theta_d)}, \nonumber
\end{eqnarray}
\end{small}
where $\mathcal{S}$ is previously defined encoder state,
$R(A,u)$ is the delayed reward \cite{yu2017seqgan} obtained after the whole card $A$ is generated
and is estimated by the following equation:
\begin{small}
\begin{equation}
\label{eq:reward}
R(A,u) = 2\times\big(P(r=1|A,u;\phi^*)-0.5\big),
\end{equation}
\end{small}
here we rescale the value of reward between $-1.0$ to $1.0$.

One problem for training REINFORCE policy is that the reward is delayed and sparse,
in which policy may be hard to receive positive reward signal, thus the training procedure of policy becomes unstable and falls into local minimal finally.
In order to effectively avoid non-optimal local minimal and steadily increase the reward throughout training,
we borrow the idea of Hill Climbing (HC) which is heuristic search used for mathematical optimization problems \cite{duan2018multi}. 
Instead of directly sampling from the policy by $A\sim P(A|S,u;\theta)$,
in our method we first stochastically sample a buffer of $m=5$ solutions (cards) from policy and select the best one as $A^*$,
then train the policy by $A^*$ according to Eq. \ref{eq:L_R}.
In that case, we will always learn from a better solution to maximize reward, 
train on it and use the trained new policy to generate a better one. 

\subsubsection{Combination}
\label{sec:loss_combination}
To benefit from both fields of \emph{Learning from Demonstrations} and \emph{Learning from Rewards},
we simply apply linear combination of their loss functions
and conduct the final loss as:
\begin{small}
\begin{equation}
\label{eq:total_loss}
\mathcal{L}(\theta) = \alpha\times \mathcal{L}_{S}(\theta) + (1-\alpha)\times \mathcal{L}_{R}(\theta),
\end{equation}
\end{small}
where $\mathcal{L}_{S}(\theta)$ and $\mathcal{L}_{R}(\theta)$ are formulated by Eq. \ref{eq:L_S} and \ref{eq:L_R},
$\alpha\in[0,1]$ is the hyper-parameter which should be tuned.
The overall learning process is shown in Algorithm \ref{alg:RLfD}.
\section{Experiments}
\label{sec:experiments}
In this section, we conduct experiments with the aim of
answering the following research questions:
\begin{enumerate}
\item[\textbf{RQ1}] Does our proposed \emph{GAttN with RLfD} method outperform the baseline methods in exact-K recommendation problem?
\item[\textbf{RQ2}] How does our proposed \emph{Graph Attention Networks} (GAttN) framework work for modeling the problem?
\item[\textbf{RQ3}] How does our proposed optimization framework \emph{Reinforcement Learning from Demonstrations} (RLfD) work for training the model?
\end{enumerate}
\subsection{Experimental Settings}
\subsubsection{Datasets}
We experiment with three datasets and Tab. \ref{tab:dataset_statistic} summarizes the statistics.
The first two datasets are constructed from MovieLens
and last dataset is collected from real-world exact-K recommendation system on Taobao platform.
The implementation details and parameter settings can be found in Appx. \ref{sec:implementation}.

\paragraph{\textbf{MovieLens.}}
This movie rating dataset\footnote{\url{http://grouplens.org/datasets/movielens/100k/}} has been
widely used to evaluate collaborative filtering algorithms in RS.
As there is no public datasets to tackle exact-K recommendation problem,
we construct for it based on MovieLens.
While MovieLens is an explicit feedback data,
we first transform it into implicit
data, where we regard the 5-star ratings as positive feedback and treat all other entries as unknown feedback \cite{wang2017irgan}.
Then we construct recommended cards of each user with set of $K$ items\footnote{The $K$ items in a card are randomly permuted. As we suppose in Sec. \ref{sec:problem_definition}, the permutation of the $K$ item is not considered.},
where cards containing positive item are regarded as positive cards (labeled as 1) and cards without any positive item are negative cards (labeled as 0).
Meanwhile, positive item in the corresponding card can be seen as what user actually clicked or preferred item belonging to that card.
Finally we construct a candidate set with $N$ items for each card for a user, 
where the $N$ items are randomly sampled from the whole items set given this user and must include all the items in that card.
We show examples how the dataset like in Tab. \ref{tab:dataset_example}.
Specially in our experiments, we construct two dataset:
1) card with $K=4$ items along with $N=20$ candidate items and 
2) card with $K=10$ items along with $N=50$ candidate items.
We call the above two dataset as \textbf{MovieLens(K=4,N=20)} and \textbf{MovieLens(K=10,N=50)}.
Notice that there is no constraint between items in the output card (i.e $C=\emptyset$ defined in Sec. \ref{sec:problem_definition}) for these two datasets.

\paragraph{\textbf{Taobao.}}
Above two datasets for exact-K recommendation problem based on MovieLens are what we call synthetic data which are not real-world datasets in production.
On the contrary, we collect a dataset from exact-K recommendation system in Taobao platform,
of which two days' samples are used for training and samples of the following day for testing,
and specifically with $K=4$ and $N=50$.
We call it \textbf{Taobao(K=4,N=50)}.
Notice there is a required constraint between items in the output card in this dataset,
that normalized edit distance (NED) \cite{marzal1993computation} of any two items' titles must be larger than $\tau=0.5$, i.e $C=\{NED(a_i,a_j)\geq\tau|a_i\in A,a_j\in A,i\neq j\}$ defined in Sec. \ref{sec:problem_definition},
to guarantee the diversity of items in a card.

\subsubsection{Evaluation Protocol}
For evaluation, we can't use traditional ranking evaluation metrics such as nDCG, MAP, etc. 
These metrics either require prediction scores for individual items or assume that ``better'' items should appear in top ranking positions,
which are not suitable for exact-K recommendation problem.
\paragraph{\textbf{Hit Ratio}} 
Hit Ratio (HR) is a recall-based metric, measuring how much the
testing ground-truth $K$ items of card $A^*$ are in the predicted card $A$ with exact $K$ items.
Specially for exact-K recommendation, we refer to HR@K and is formulated as follows:
\begin{small}
\begin{equation}
HR@K = \left.\sum_{i=1}^{n}{\frac{|A_i\cap A_i^*|}{K}}\middle/n\right.,
\end{equation}
\end{small}
where $n$ is the number of testing samples,
$|\cdot|$ represents the number of items in a set.

\paragraph{\textbf{Precision}}
Precision (P) measures whether the actually clicked (positive) item $a^*$ in ground-truth card is also included in the predicted card $A$ with exact $K$ items,
and is formulated as follows:
\begin{small}
\begin{equation}
P@K = \left.\sum_{i=1}^{n}{I(a_i^*\in A_i)}\middle/n\right.,
\end{equation}
\end{small}
where $I(\cdot)\in\{0,1\}$ is the indicator function. 

\subsubsection{Baselines}
Our baseline methods are based on Naive Node-Weight Estimation Method (in Sec. \ref{sec:nnwem}) to adapt to exact-K recommendation.
The center part is to estimate node weight which can be seen as CTR for the corresponding item.
Therefor LTR based methods are applied and we compare with the follows:
\paragraph{\textbf{Pointwise Model.}}
DeepRank model is a popular ranking method in production which applies DNNs and a pointwise ranking loss (a.k.a MLE) \cite{he2017neural}.

\paragraph{\textbf{Pairwise Model.}}
BPR \cite{rendle2009bpr} is the method optimizes MF model \cite{koren2009matrix} with a pairwise ranking loss. It is a highly competitive
baseline for item recommendation.

\paragraph{\textbf{Listwise Model.}}
GRU based listwise model (Listwise-GRU) a.k.a DLCM \cite{ai2018learning} is a SOTA model for whole-page ranking refinement.
It applies GRU to encode the candidate items with a list-wise ranking loss.
In addition, we also compare with listwise model based on Multi-head Self-attention in Sec. \ref{sec:encoder} as Listwise-MHSA.

\subsection{Performance Comparison (RQ1)}
Tab. \ref{tab:main_result} shows the performances of P@K and HR@K for the three datasets with respect to different methods.
First, we can see that our method with the best setting (\emph{GAttN with RLfD}) achieves the best performances
on both datasets, significantly outperforming the SOTA
methods Listwise-MHSA and Listwise-GRU by a large margin
(on average over three datasets,
the relative improvements for P@K and HR@K are
\textbf{7.7\%} and \textbf{4.7\%}, respectively).
Secondly from the results, 
we can find that listwise methods (both Listwise-MHSA and Listwise-GRU)
outperform pointwise and pariwise baselines significantly.
Therefore listwise methods are more suitable for exact-K recommendation,
because they consider the context information to represent an item (node in graph) as what we have claimed in Sec. \ref{sec:encoder}.
And Listwise-MHSA performs better than Listwise-GRU,
which indicates the effectiveness of our proposed MHSA method for encoding the candidate items (graph nodes).
More detailed analysis for our method \emph{GAttN with RLfD} can be found in the following two subsections (RQ2 and RQ3).
\begin{table}[h]
	\caption{Overall performances respect to different methods on three datasets, where $*$ means a statistically significant improvement for $p<0.01$.}
	\centering
	\scriptsize
	\begin{tabular}{c|c|c|c|c|c|c}
		\toprule
		\multirow{2}{*}{Model} & \multicolumn{2}{c|}{MovieLens (K=4,N=20)} & \multicolumn{2}{c|}{MovieLens (K=10,N=50)} & \multicolumn{2}{c}{Taobao (K=4,N=50)} \\
		\cmidrule{2-7}
		 & P@4 & HR@4 & P@10 & HR@10 & P@4 & HR@4 \\
		 \cmidrule{1-7}
		 DeepRank & 0.2120 & 0.1670 & 0.0854 & 0.1320 & 0.6857 & 0.6045 \\
		 BPR &  0.3040 & 0.2050 & 0.2350 & 0.1801 & 0.7357 & 0.6582 \\
		 Listwise-GRU & 0.4142 & 0.2423 & 0.4041 & 0.2144 & 0.7645 & 0.6942 \\
		 Listwise-MHSA & 0.4272 & 0.2465 & 0.4384 & 0.2168 & 0.7789 & 0.7176 \\
		 \textbf{Ours (best)} & \textbf{0.4743} & \textbf{0.2611} & \textbf{0.4815} & \textbf{0.2245} & \textbf{0.7958} & \textbf{0.7488} \\
		 \cmidrule{1-7}
		 Impv. & 11.0\%$^*$ & 6.1\%$^*$ & 9.8\%$^*$ & 3.6\% & 2.2\% & 4.3\% \\
		\bottomrule
	\end{tabular}
	\label{tab:main_result}
\end{table}

\subsection{Analysis for GAttN (RQ2)}
\begin{figure}[th]
	\centering
	\includegraphics[angle=0, width=0.7\columnwidth]{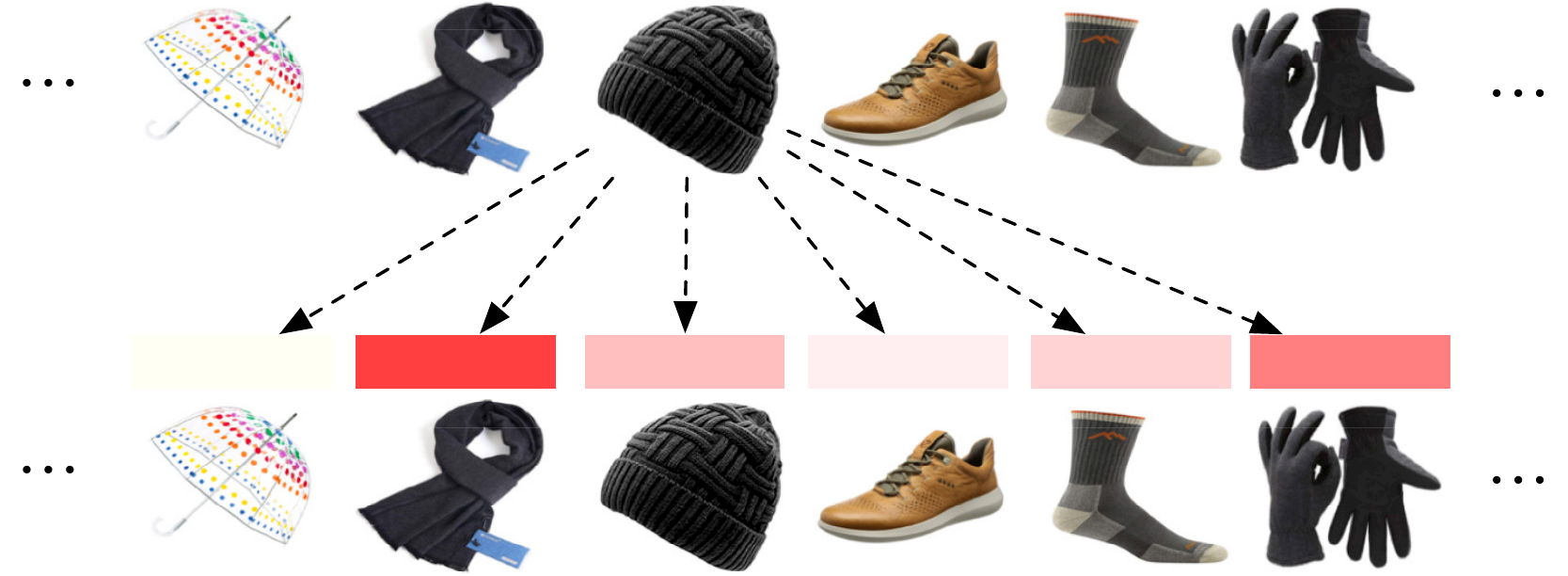}
	\caption{An example of the attention mechanism in the encoder self-attention in layer 2.
	The higher attention weights of the item, the darker color of the grid.
	We take item ``hat'' for example and only show attention weights in one head.}
	\label{fig:attention}
\end{figure}

Tab. \ref{tab:main_result} shows that Listwise-MHSA performs better than Listwise-GRU
on both P@K and HR@K on all datasets.
It indicates the effectiveness to apply MHSA method for encoding the candidate items (graph nodes).
As we claimed in Sec. \ref{sec:encoder}, 
the representation for a node should consider the other nodes in graph, 
for there can exist some underlying structures in graph that nodes may influence between each other.
Here we further give a presentational case on how the self-attention works in encoder (see Fig. \ref{fig:attention}) based on Taobao dataset.
Take item ``hat'' in graph for example, items with larger attention weights to it are kinds of ``scarf'', ``glove'' and ``hat''.
It is reasonable that users focusing on ``hat'' tend to prefer ``scarf'' rather than ``umbrella''.
So to represent item ``hat'' in graph, it's helpful to attend more features of items like ``scarf''.

\begin{figure}[th]
	\centering
	\includegraphics[angle=0, width=0.84\columnwidth]{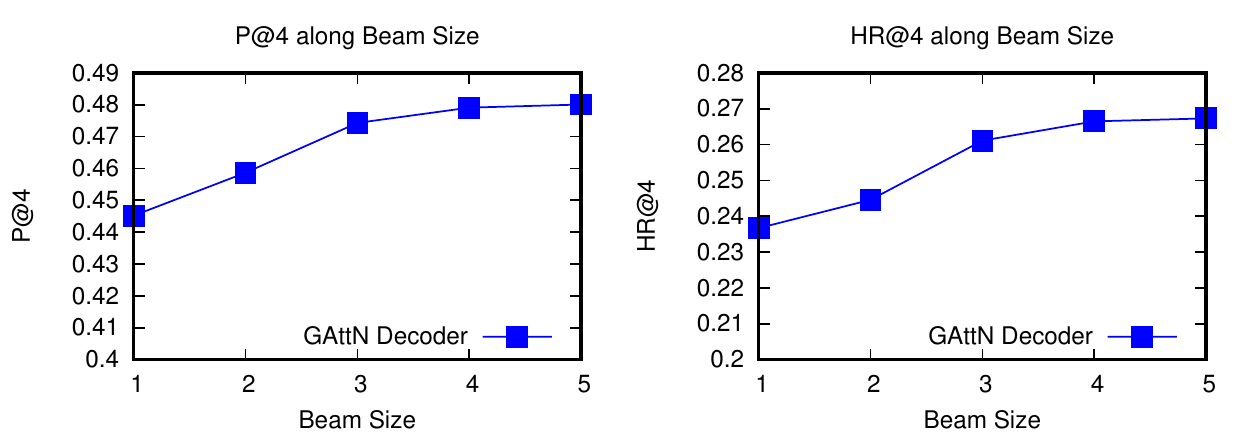}
	\caption{Performance of P@4 and HR@4 with different beam size in GAttN decoder.}
	\label{fig:beam_search}
\end{figure}

Beam search is proposed in GAttN decoder (see Sec. \ref{sec:encoder}) to expand the search space and try more combinations of items in a card to get a most optimal solution.
A critical hyper-parameter for beam search is the beam size, indicating how many solutions to search in a decoding time-step.
We tune beam size in Fig. \ref{fig:beam_search} and find that larger beam size can lead to better performances\footnote{We only report the results on MovieLens(K=4,N=20) and other datasets follow the same conclusion.} on both P@K and HR@K.
However we can also see that when beam size gets larger than 3 the improvement of performances will be minor,
so for efficiency consideration we set beam size as 3 in our experiments.

\subsection{Analysis for RLfD (RQ3)}
To verify how our proposed optimization framework \emph{RLfD} works,
we will do the following ablation tests:
\begin{enumerate}[(1)]
\item Set $\alpha=0$ in Eq. \ref{eq:total_loss} (only reinforcement loss as Eq. \ref{eq:L_R}), and compare \emph{RL(w/ hill-climbing)} with \emph{RL(w/o hill-climbing)}.
\item Set $\alpha=1$ in Eq. \ref{eq:total_loss} (only supervised loss as Eq. \ref{eq:L_S}), and compare \emph{SL(w/ policy-sampling)} with \emph{SL(w/o policy-sampling)}.
\item Finetune $\alpha$ in Eq. \ref{eq:total_loss} and figure out the influence to the combination of SL with RL.
\end{enumerate}
Here we represent \emph{Learning from Demonstrations} and \emph{Learning from Rewards} as SL and RL for short.
Tab. \ref{tab:ablation_test} gives an overall results and we only report on dataset MovieLens(K=4,N=20) for simplification.
\begin{table}[h]
	\caption{Performance for different settings in RLfD.}
	\centering
	\scriptsize
	\begin{tabular}{cc|c|c}
		\toprule
		\multirow{2}{*}{} & \multirow{2}{*}{Settings in RLfD} & \multicolumn{2}{c}{MovieLens (K=4,N=20)} \\
		\cmidrule{3-4}
		& & P@4 & HR@4 \\
		\cmidrule{1-4}
		1 & RL(w/o hill-climbing) & 0.3340 & 0.2314 \\
		2 & RL(w/ hill-climbing) &  0.3573 & 0.2330 \\
		3 & SL(w/o policy-sampling) & 0.4095 & 0.2401 \\
		4 & SL(w/ policy-sampling) & 0.4272 & 0.2465 \\
		5 & RL(w/o hill-climbing) + SL(w/ policy-sampling) & 0.4495 & 0.2514 \\
		6 & RL(w/ hill-climbing) + SL(w/o policy-sampling) & 0.4472 & 0.2534 \\
		7 & \textbf{RL(w/ hill-climbing) + SL(w/ policy-sampling)} & \textbf{0.4743} & \textbf{0.2611} \\
		\bottomrule
	\end{tabular}
	\label{tab:ablation_test}
\end{table}

\begin{figure}[th]
	\centering
	\includegraphics[angle=0, width=0.84\columnwidth]{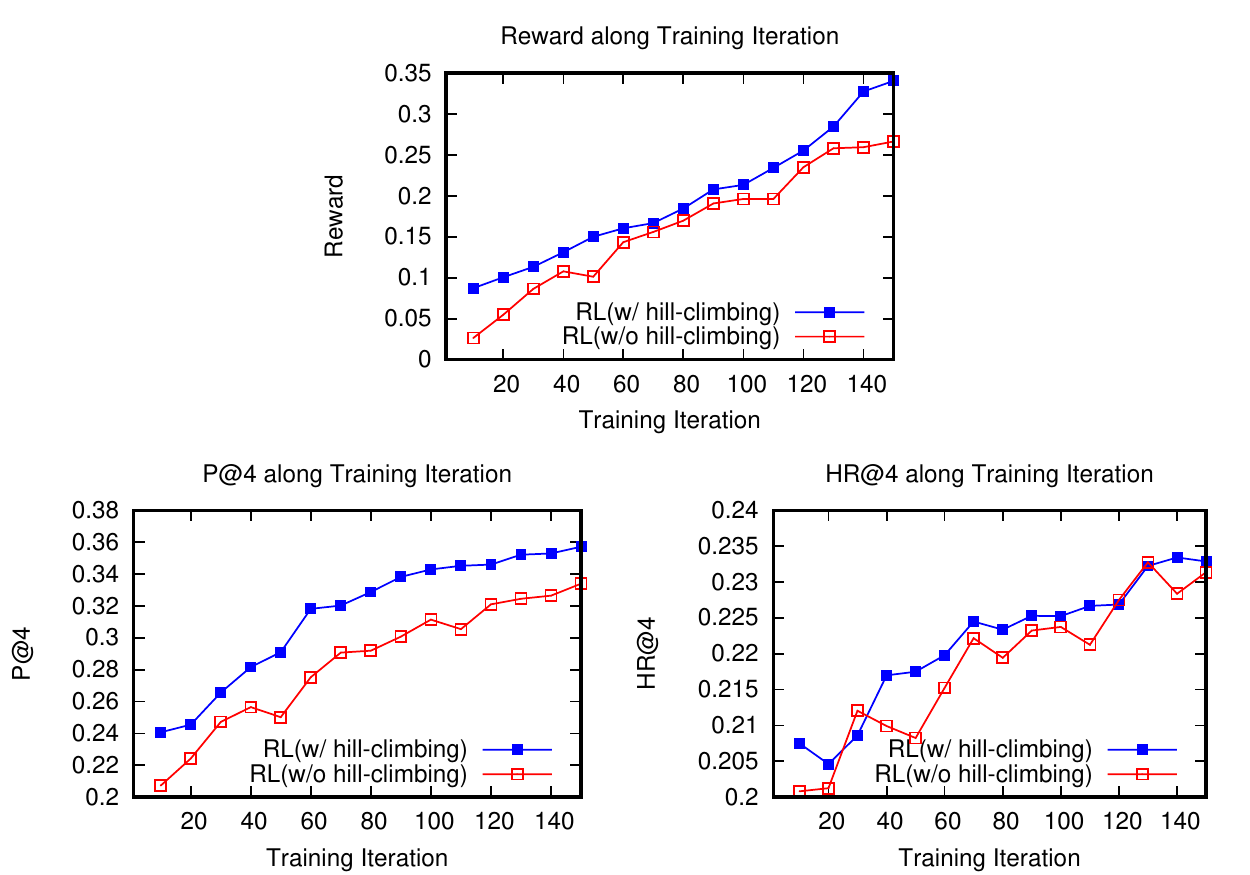}
	\caption{Learning curves respect to Reward, P@4 and HR@4 for RL with (w/) or without (w/o) hill-climbing.}
	\label{fig:hill_climbing}
\end{figure}

Fig. \ref{fig:hill_climbing} shows the learning curves respect to Reward (defined in Eq. \ref{eq:reward}), P@4 and HR@4 for RL with (w/) or without (w/o) hill-climbing proposed in Sec. \ref{sec:reinforce}.
From the curves, we can find that with the help of hill-climbing REINFORCE training becomes more stable and steadily improves the performance, 
finally achieves a better solution (row 1 vs. 2 and row 5 vs. 7 in Tab. \ref{tab:ablation_test}).
Another insight in Fig. \ref{fig:hill_climbing} is that learning curve of Reward is synchronous monotonous with P@4 and HR@4,
which verifies the effectiveness of our defined reward function to direct the objective in problem.

\begin{figure}[th]
	\centering
	\includegraphics[angle=0, width=0.84\columnwidth]{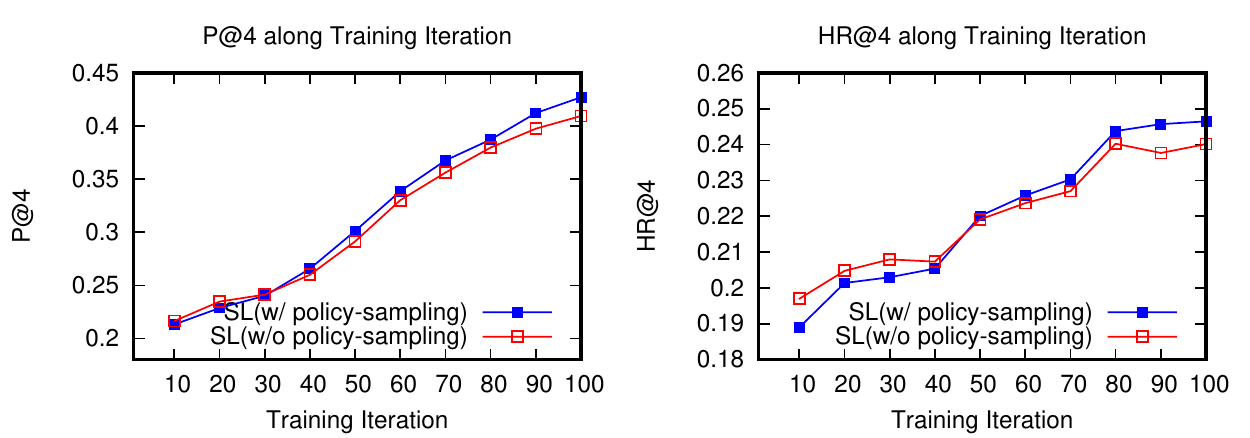}
	\caption{Learning curves respect to P@4 and HR@4 for SL with (w/) or without (w/o) policy-sampling.}
	\label{fig:policy_sampling}
\end{figure}

Fig. \ref{fig:policy_sampling} shows the learning curves respect to P@4 and HR@4 for SL with (w/) or without (w/o) policy-sampling proposed in Sec. \ref{sec:supervise}.
We observe that in the beginning 50 iterations SL with policy-sampling may perform worse than without policy-sampling.
We believe that in the first steps of training procedure the learned policy can be poor,
so feeding the output sampled from such policy to the next time-step as input in decoder can lead to worse performances.
However as training goes on, SL with policy-sampling will converge better for revising the inconsistency between training and inference of policy,
finally achieve better performances (row 3 vs. 4 and row 6 vs. 7 in Tab. \ref{tab:ablation_test}).

\begin{figure}[th]
	\centering
	\includegraphics[angle=0, width=0.84\columnwidth]{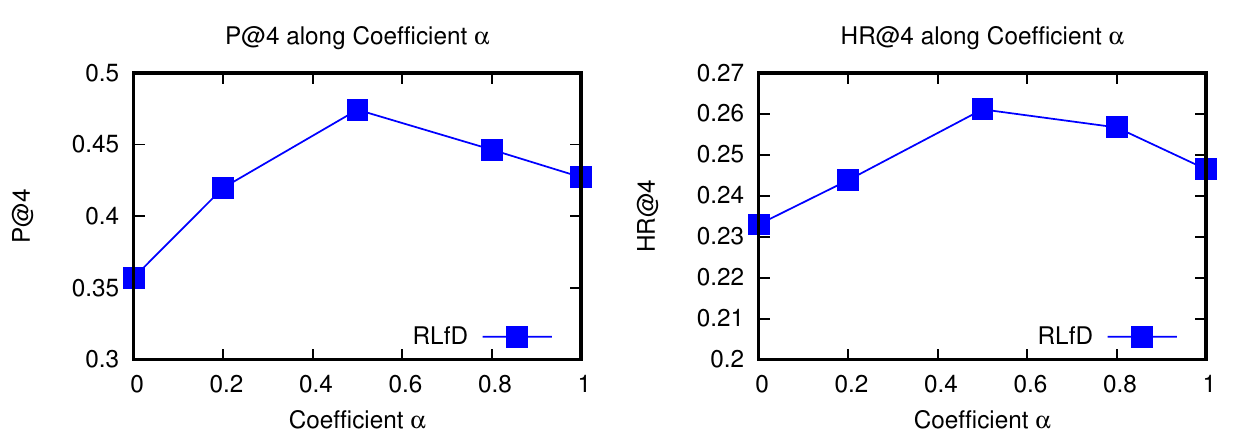}
	\caption{Performance of P@4 and HR@4 with different coefficients $\alpha$ in loss defined in Eq. \ref{eq:total_loss}.}
	\label{fig:coefficient}
\end{figure}

In Fig. \ref{fig:coefficient} we tune hyper-parameter $\alpha$ defined in Eq. \ref{eq:total_loss},
which represents trade-off for applying SL and RL in training process.
We observe that $\alpha=0.5$ achieves the best both on P@4 and HR@4.
The performances increase when
$\alpha$ is tuned from 0 to the optimal value and then
drops down afterwards,
which indicates that properly combining SL and RL losses can result in the best solution.
Furthermore, we find that when only apply SL loss ($\alpha=1$) we can get a preliminary sub-optimal policy,
after involving some degree of RL loss the policy can be directed to achieve more optimal solutions,
which verifies the sufficiency-and-efficiency of our proposed \emph{Reinforcement Learning from Demonstrations} to train the policy.
\section{Conclusion and Future Work}
This work targets to a practical recommendation problem named exact-K recommendation,
we prove that it is different from traditional top-K recommendation.
In the first step, we give a formal problem definition,
then reduce it to a Maximal Clique Optimization problem which is a combinatorial optimization problem and NP-hard.
To tackle this specific problem, we propose a novel approach of \emph{GAttN with RLfD}.
In our evaluation, we perform extensive analysis to demonstrate the highly positive effect of our proposed method targeting exact-K recommendation problem.
In our future work, we plan to adopt adversarial training for the components of Reward Estimator and REINFORCE learning, regarding as discriminator and generator in GAN's \cite{wang2017irgan} perspective.
Moreover further online A/B testing in production will be conducted.

\bibliographystyle{ACM-Reference-Format}
\balance
\bibliography{ref}

\appendix
\section{Naive Node-Weight Estimation Method}
\begin{algorithm}
	\caption{Naive Node-Weight Estimation Method.}       
	\label{alg:naive} 
	\begin{algorithmic}[1]
		\Require Given user $u$ and candidate items set $S$, construct graph $\mathbb{G}(\mathcal{N},\mathcal{E})$ defined in paragraph 2 of Sec. \ref{sec:problem_definition}.
		\State Estimate weight $w_i$ of each node $n_i\in \mathcal{N}$ in graph based on CTR of corresponding item $s_i\in S$.
		\State Initial result card $A=\emptyset$.
		\For {\emph{t = 1 to K}}
		\State Select node $a_t$ with the largest weight in $\mathcal{N}$ and add to $A$.
		\State Remove $a_t$ and nodes in $\mathcal{N}$ which are not adjacent to $a_t$.
		\EndFor\\
		\Return result card $A$.
	\end{algorithmic}
\end{algorithm}

\section{Reinforcement Learning from Demonstrations}
\begin{algorithm}
	\caption{Reinforcement Learning from Demonstrations.}       
	\label{alg:RLfD}
	\begin{algorithmic}[1]
		\Algphase{Phase 1 - Reward Estimator Training}
		\Require reward function $P(r=1|A,u;\phi)$, dataset $P_{data}^{D}(r^*|A,u)$
		\State Optimize $\phi$ with gradient descent by loss function $\mathcal{L}_{D}(\phi)$.\\
		\Return $P(r=1|A,u;\phi^*)$
	\end{algorithmic}
	\begin{algorithmic}[1]
		\Algphase{Phase 2 - Policy Training}
		\Require optimized reward function $P(r=1|A,u;\phi^*)$, dataset $P_{data}^{S}(A^*|S,u)$, policy $P(A|S,u;\theta)$
		\State Optimize $\theta$ with gradient descent by loss function $\mathcal{L}(\theta)$.\\
		\Return $P(A|S,u;\theta^*)$
	\end{algorithmic}
\end{algorithm}

\section{Experimental Settings}
\subsection{Datasets}
\begin{table}[h]
	\caption{Statistics of the experimented datasets.}
	\label{tab:dataset_statistic}
	\centering
	\scriptsize
	\begin{tabular}{c|c|c|c|c}
		\toprule
		\textbf{Dataset} & \textbf{User\#} & \textbf{Card\#} & \textbf{Item\#} & \textbf{Sample\#} \\
		\midrule
		MovieLens(K=4,N=20) & 817 & 40036 & 1630 & 40036 \\
		\midrule
		MovieLens(K=10,N=50) & 485 & 33196 & 1649 & 33198 \\
		\midrule
		Taobao(K=4,N=50) & 581055 & 310509 & 3148550  & 1116582 \\
		\bottomrule
	\end{tabular}
\end{table}

\begin{table}[h]
	\caption{Show case of the dataset.}
	\label{tab:dataset_example}
	\centering
	\scriptsize
	\begin{threeparttable}
	\begin{tabular}{c|c|c|c|c|c}
		\toprule
		& \textbf{user} & \textbf{card} & \textbf{candidate items} & \textbf{card label} & \textbf{positive item} \\
		\midrule
		sample\#1 & 1 & 1,2,3,4 & 1,2,3,4,...,20 & 1 & 2 \\
		sample\#2 & 1 & 1,4,5,6 & 1,2,3,4,...,20 & 0 & / \\
		\multicolumn{6}{c}{...} \\
		\bottomrule
	\end{tabular}
	\begin{tablenotes}
		\footnotesize
		\item (We take $K=4$ and $N=20$ for example. Items and users are represented as IDs here. Card label represents whether the card is clicked or satisfied by user (labeled as 1) or not (labeled as 0). Positive item is the actually clicked item in card by user.)
	\end{tablenotes}
	\end{threeparttable}
\end{table}

\subsection{Implementation and Parameter Settings}
\label{sec:implementation}
Here we report implementation details for the three datasets\footnote{\url{https://github.com/pangolulu/exact-k-recommendation}} (two MovieLens based datasets and one Taobao based dataset),
and our implementation is based on TensorFlow\footnote{\url{https://www.tensorflow.org/}}.
To construct the training and test sets, we perform a 4:1 random splitting as in \cite{wang2017irgan} for all the datasets.
\subsubsection{MovieLens}
\label{sec:implementation_movielens}
Notice both MovieLens(K=4,N=20) and MovieLens(K=10,N=50) share the same parameter settings.
For a fair comparison, all models are set with an embedding size of $16$ for item and user IDs,
and optimized using the mini-batch Adam optimizer with a batch size of $32$ and learning rate of $0.001$.
All models are trained for $10$ epoch.
All the trainable feed-forward parameter matrices are set with the same input and output dimension as $32\times32$ (including DeepRank, BPR, and all the RNN cells in both Listwise-GRU, Listwise-MHSA and ours). 
Specifically for our GAttN model, in decoder (in Sec. \ref{sec:decoder}) we use LSTM cells with units number of 32 and set beam size as $3$, number of heads in encoder (in Sec. \ref{sec:encoder}) MHSA layer is $2$, and the coefficient parameter $\alpha$ in loss function (in Sec. \ref{sec:loss_combination}) is $0.5$.
Number of layers $L$ in both encoder and decoder are set as 2.
For reward estimator model (in Sec. \ref{sec:reinforce}), we set the hidden size in fully-connected layer as 128.

\subsubsection{Taobao}
\label{sec:implementation_taobao}
In this dataset, the feature vectors for user and item are statistic features with size of 40 and 52 specifically, instead of ID features.
Sample statistic features are PV (page view), IPV (item page view), GMV (cross merchandise volume), CTR (click through rate) and CVR (conversion rate) for 1 day, 7 days and 14 days, etc. 
For this dataset, we first transfer the input representation of user and item to 32 dimension, i.e we set $W_I\in \mathbb{R}^{92\times32}$ and $b_I\in \mathbb{R}^{32}$ in Sec. \ref{sec:input}.
And all the other hyper-parameters are set as the same with those on MovieLens based datasets (refer to Appx. \ref{sec:implementation_movielens}).

\end{document}